\documentclass[twocolumn]{aastex62}

\graphicspath{{./}{figures/}}
\usepackage[utf8]{inputenc}
\usepackage{graphicx}% Include figure file
\usepackage{dcolumn}% Align table columns on decimal point
\usepackage{bm}% bold math
\usepackage{ulem} %here for \sout, removed to avoid funny underlines
\usepackage{hyperref}% add hypertext capabilities
\usepackage{multirow,enumitem}
\usepackage{amsmath}
\usepackage{color}
\usepackage{xcolor}
\usepackage{multirow}

\newcommand{\Om}{\Omega_m}
\newcommand{\LCDM}{\rm{\Lambda}CDM}

\begin{document}
\title{A Simple Phenomenological Emergent Dark Energy Model can Resolve the Hubble Tension}
\author{Xiaolei Li}
\affiliation{Korea Astronomy and Space Science Institute, Daejeon 34055, Korea}
\affiliation{Quantum Universe Center, Korean Institute of Advanced Studies, Hoegiro 87, Dongdaemun-gu, Seoul 130-722, Korea}
\author{Arman Shafieloo}
\affiliation{Korea Astronomy and Space Science Institute, Daejeon 34055, Korea}
\affiliation{University of Science and Technology, Yuseong-gu 217 Gajeong-ro, Daejeon 34113, Korea}
\date{\today}% It is always \today, today,
 
\begin{abstract}
Motivated by the current status of the cosmological observations and significant tensions in the estimated values of some key parameters assuming the standard $\Lambda$CDM model, we propose a simple but radical phenomenological emergent dark energy model where dark energy has no effective presence in the past and emerges at the later times. Theoretically, in this phenomenological dark energy model with zero degree of freedom (similar to a $\Lambda$CDM model), one can derive that the equation of state of dark energy increases from $-\frac{2}{3 {\rm{ln}}\, 10} -1$ in the past to $-1$ in the future. We show that by setting a hard-cut 2$\sigma$ lower bound prior for the $H_0$ that associates with $97.72\%$ probability from the recent local observations~\citep{riess2019large}, this model can satisfy different combinations of cosmological observations at low and high redshifts (SNe Ia, BAO, Ly{$\alpha$} BAO and CMB) substantially better than the concordance $\Lambda$CDM model with $\Delta \chi^2_{bf} \sim -41.08$ and $\Delta\,{\rm{DIC}} \sim-35.38$.
If there are no substantial systematics in SN Ia, BAO or Planck CMB data and assuming reliability of the current local $H_0$ measurements, there is a very high probability that with slightly more precise measurement of the Hubble constant our proposed phenomenological model rules out the cosmological constant with decisive statistical significance and is a strong alternative to explain combination of different cosmological observations. This simple phenomenologically emergent dark energy model can guide theoretically motivated dark energy model building activities. 

\end{abstract}

\section{Introduction}
While current cosmological observations have been in great agreement with the standard $\LCDM$ model, 
there is significant tensions of some key cosmological parameters derived by assuming this model.
One of the major issues is the inconsistency between the local measurement of the Hubble constant by \textit{the Supernova H0 for the Equation of State(SH0ES)} collaboration \citep{riess20162,riess2018type,riess2019large} and the estimation of this parameter using \textit{Planck} cosmic microwave background (CMB) and other cosmological observations assuming $\LCDM$ model \citep{ade2016planck,aghanim2018planck}. Another issue is  the estimation of the $\Omega_{m}h_{0}^2$ from the baryon acoustic oscillation (BAO) measurement at $z\,=\,2.34$ from BOSS and eBOSS surveys using Ly$\alpha$ forest and the estimated values from \textit{Planck} CMB observations assuming $\LCDM$ model \citep{sahni2014model,ding2015there,zheng2016omh,sola2017h0,alam2017constraining,shanks2018gaia}. 

A possible solution to this issue may be a carefully constructed yet simple alternative model of dark energy that can satisfy all of the observations, or an unconventional model of the early Universe \citep{hazra2019parameter}. 

In this letter we propose a simple (zero degree of freedom) but radical phenomenological model of dark energy with symmetrical behavior around the current time where dark energy and matter densities are comparable. In this model dark energy has no effective presence in the past and emerges at later times. Setting hard-cut priors from local measurements of the Hubble constant , we confront this model with combination of low and high redshift cosmological observations, namely SNe Ia data, BAO data (including BAO Ly$\alpha$ measurement) and CMB measurement and show that significantly it can outperform statistically the standard $\Lambda$CDM model as well as the $w_0$-$w_a$ parameterization. 

This letter is organised as follows: in section~\ref{sec:cos_model} we briefly introduce the Friedmann equations for our model. The observational data to be used, including SNe Ia, BAO and distance prior from CMB, are presented in section \ref{sec:analysis}. Section~\ref{sec:res} contains our main results and some discussion. We conclude in section~\ref{sec:con}.

\section{Phenomenologically Emergent Dark Energy Model (PEDE)} \label{sec:cos_model}
The Hubble parameter within the Friedmann-Lema\^{\i}tre-Robertson-Walker (FLRW) metric, assuming a flat universe, could be described as:
\begin{equation}
    H^2(z)\,=\,H_0^2\left[{\Om} (1+z)^3+ \widetilde{\Omega}_{\rm{DE}}(z)   \right]
\end{equation}
where $\Om$ is the matter density at present time and $\widetilde{\Omega}_{\rm{DE}}(z)$ can be expressed as:
\begin{equation}
    \widetilde{\Omega}_{\rm{DE}}(z)\,=\,\Omega_{\rm{DE,0}}\times {\rm{exp}} \left[3\int_0^z \frac{1+w(z')}{1+z'} dz' \right]
\end{equation}
where $w(z)\,=\,p_{\rm{DE}}/\rho_{\rm{DE}}$ is the equation of state of Dark Energy.

In $\LCDM$ model, $w(z) = -1$ and $\widetilde{\Omega}_{\rm{DE}}(z)\,= (1-\Om)\,=\,$ {{constant}}. 
%For wCDM, $w(z)\,=\,constant$ and $\Ode\,=\,\Omega_{\rm{DE,0}}(1+z)^{3(1+w)}$. 
For the widely used CPL parameterization model ($w_0$-$w_a$ model) \citep{chevallier2001accelerating,linder2003cosmic}, the equation of state of dark energy is given by $w(z)\,=\,w_0+\frac{w_a z}{1+z}$ so one can derive $\widetilde{\Omega}_{\rm{DE}}(z)\,=\,\Omega_{\rm{DE,0}}(1+z)^{3(1+w_0+w_a)}{\rm{exp}}(\frac{-3w_a z}{1+z})$. 

In this letter, we introduce the PEDE model in which the dark energy density has the following form:
\begin{equation}
\widetilde{\Omega}_{\rm{DE}}(z)\,=\,\Omega_{\rm{DE,0}}\times \left[ 1 - {\rm{tanh}}\left( {\rm{log}}_{10}(1+z) \right) \right] 
\end{equation}
where $\Omega_{\rm{DE,0}}=1-\Omega_{0m}$ and $1+z=1/a$ where $a$ is the scale factor. This dark energy model has no degree of freedom (similar to the case of $\Lambda$CDM model) and we can derive its equation of state following:

\begin{equation}
    w(z) \,=\,\frac{1}{3} \frac{d\,{\rm{ln}}\, \widetilde{\Omega}_{\rm{DE}}}{d z} (1+z)-1
\end{equation}

where we get, 

\begin{align}
    w(z)& \,=\,-\frac{1}{3 {\rm{ln}}\, 10} \times \frac{1-{\rm{tanh}}^2\left[{\rm{log}}_{10}(1+z)\right]}{1-{\rm{tanh}}\left[{\rm{log}}_{10}\,(1+z)\right]} -1 \\
        & \,=\,-\frac{1}{3 {\rm{ln}}\, 10} \times\left({1+{\rm{tanh}}\left[{\rm{log}}_{10}\,(1+z)\right]}\right) -1. 
\end{align}

\begin{figure}[h!]
\centering
\includegraphics[width=0.45\textwidth]{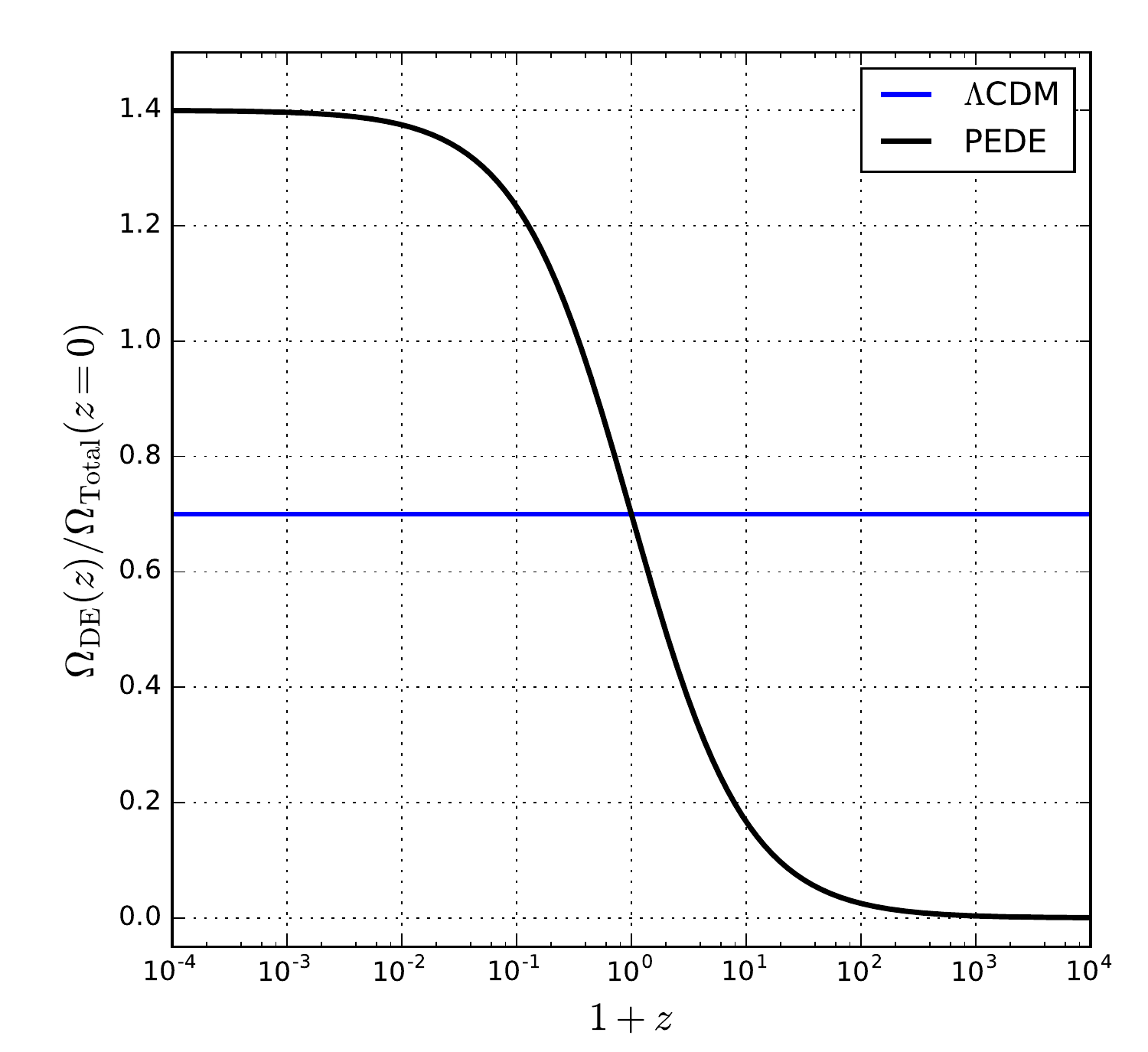}
\includegraphics[width=0.45\textwidth]{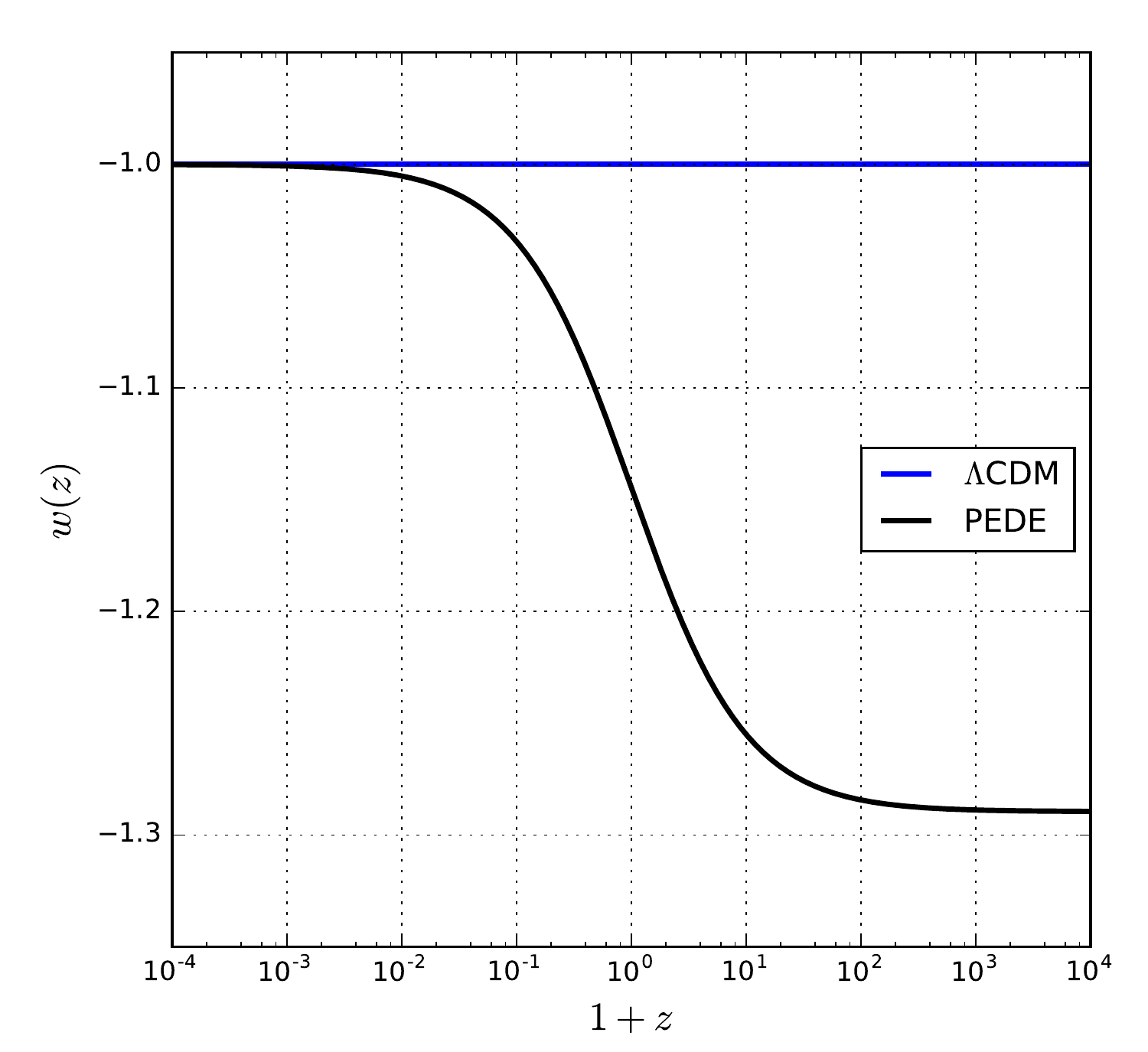}
\caption{The upper plot shows the evolution of dark energy density $\Omega_{\rm{DE}}(z)$ from early times to the far future and the bottom plot presents the evolution of Equation of State of Dark Energy $w(z)$ for $\Lambda$CDM and PEDE models. This figure is only for demonstrating the behavior of this model in comparison with cosmological constant and flatness and $\Omega_m\,=\,0.3$ is assumed for both $\LCDM$ and PEDE models.}
\label{fig:wz_ode}
\end{figure}

Note that in this model, the equation of state of dark energy at the early times would be $w(z)=-\frac{2}{3 {\rm{ln}}\, 10} -1$ and it will evolve asymptotically to $w(z)=-1$ in the far future. In this model we have $w(z=0)=-\frac{1}{3 {\rm{ln}}\, 10} -1$ at the present for the dark energy. In Fig.~\ref{fig:wz_ode}, we can see the behavior of this dark energy model in comparison to $\Lambda$. 
We should note that we can consider a more generalized form of this emergent dark energy model introducing one or more degrees of freedom such as having $\widetilde{\Omega}_{\rm{DE}}(z)\,=\,\Omega_{\rm{DE,0}}\times \frac{F(z)}{F(z=0)}$ with $F(z)\,=\,1 - {\rm{tanh}}\left(  \left[{\rm{log}}_{10}(1+z)-{\rm{log}}_{10}(1+z_t) \right] \right)$ where $z_t$ is the transition redshift (similar models have been discussed in \citet{bassett2002late,shafieloo2009cosmic}), but our results show that there is no statistical need to introduce an additional degree of freedom for this model. We can also use this generalized form and set $z_t$ to be the redshift of dark energy-matter density equality where in this case there will not be any additional degree of freedom. The behavior or this generalized form with its self-tuning characteristics will be discussed in future works. 

\section{Analysis}\label{sec:analysis}
In order to place constraints on the Dark Energy models we described above, we consider different observations in our work, including:

\begin{enumerate}[label=(\roman*)]
\item SNe Ia: we use the new "Pantheon" sample \citep{scolnic2017complete}, which is the largest combined sample of SN Ia and consists of 1048 data with redshifts in the range $0.01\,<\,z\,<\,2.3$.  In order to reduce the impact of calibration systematics on cosmology, the Pantheon compilation uses cross-calibration of the photometric systems of all the subsamples used to construct the final sample.

\item BAOs:  four lower redshift BAO data sets are used: 6-degree Field Galaxy Survey (6dFGS) \citep{beutler20116df}, the SDSS Data Release 7 Main Galaxy sample (MGS) \citep{ross2015clustering}, the BOSS DR12 galaxies \citep{alam2017clustering} and the eBOSS DR14 quasars \citep{zhao2018clustering}. In addition to these lower BAO measurement, a higher redshift BAO measurement which is derived from the cross-correlation of Ly$\alpha$ absorption and quasars in eBOSS DR14 was also used \citep{blomqvist2019baryon,agathe2019baryon}.

\item Cosmic Microwave Background: we include CMB in our analysis by using the CMB distance prior, the acoustic scale $l_{\rm{a}}$ and the shift parameter $R$ together with the baryon density $\Omega_bh^2$. The shift parameter is defined as 
\begin{equation}
R\,\equiv\,\sqrt{\Omega_mH_0^2}r(z_*)/c
\end{equation}
and the acoustic scale is 
\begin{equation}
l_a\,\equiv\,\pi r(z_*)/r_s(z_*)
\end{equation}
where $r(z_*)$ is the comoving distance to the photon-decoupling epoch $z_*$. We use the distance priors from the finally release \textit{Planck} TT, TE, EE +low E data in 2018 \citep{Chen_2019}, which makes the uncertainties 40\% smaller than those from \textit{Planck} TT+low P.
\end{enumerate}

In our analysis, we consider two kinds of data combinations. The first is Lower redshift measurements: the Pantheon supernova compilation in combination with lower redshift BAO measurements from 6dFGS, MGS, BOSS DR12 and  eBOSS DR14, hereafter we refer to as {Pantheon+BAO}. The second includes higher redshift observations from Ly$\alpha$ BAO measurements and CMB data (hereafter referred to as {Pantheon+BAO+Ly$\alpha$+CMB}.
In addition to the data combinations, 2$\sigma$ and 1$\sigma$ hard-cut $H_0$ priors, based on local measurement from \citet{riess2019large} $H_0\,=\, 74.03\pm 1.42 $ is used.

When using SNe Ia and BAO as cosmological probes, we use a conservative prior for $\Omega_{\rm{b}}h^2$ based on the measurement of D/H by \citet{cooke2018one} and standard BBN with modelling uncertainties. The constraint results are obtained with Markov Chain Monte Carlo (MCMC) estimation using \texttt{CosmoMC} \citep{lewis2002cosmological}. For quantitative comparison between our proposed model, $\Lambda$CDM model and CPL parameterization, we employ the deviance information criterion (DIC)\citep{spiegelhalter2002bayesian,liddle2007information}, defined as 
\begin{equation}
   { \rm{DIC}}\,\equiv \, D(\Bar{\theta})+2p_D\,=\,\overline{D(\theta)}+p_D,
\end{equation}\label{eq:DIC}
where
$ p_D\,=\,\overline{D(\theta)}-D(\Bar{\theta})$ and $D(\Bar{\theta})\,=\,-2\, {\rm{ln}}\, \mathcal{L}+C$, here $C$ is a 'standardizing' constant depending only on the data which will vanish from any derived quantity and $D$ is the deviance of the likelihood. If we define an effective $\chi^2$ as usual by $\chi^2\,=\,-2\,\rm{ln}\,\mathcal{L}$, we can write 
 \begin{equation}
     p_D\,=\,\overline{\chi^2(\theta)}-\chi^2(\overline{\theta}).
 \end{equation}
 
We will show that by considering the priors for the Hubble constant, our proposed model can outperform both $\Lambda$CDM model and $w_0$-$w_a$ parameterization by comparing their best fit likelihoods as well as their derived deviance  information criterion.

\section{Results} \label{sec:res}

\begin{figure*}
\centering
\includegraphics[width=0.3\textwidth]{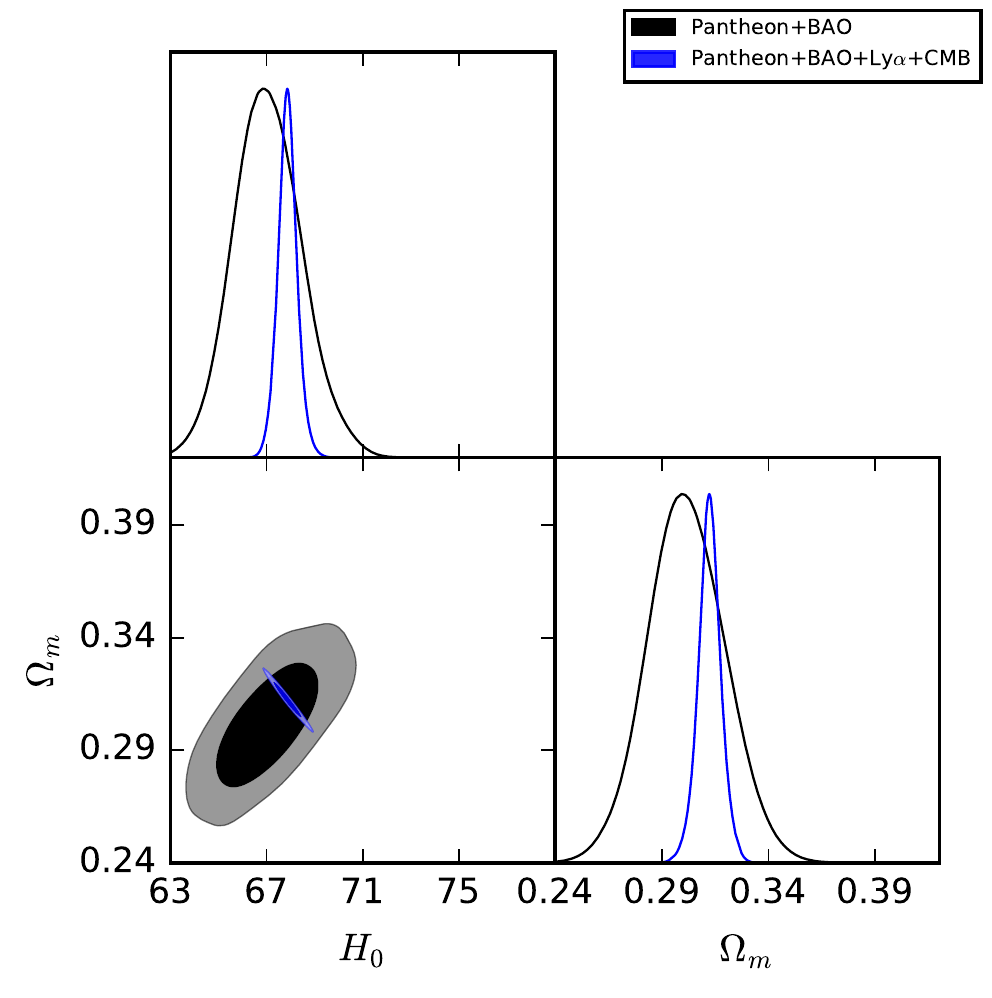} 
\includegraphics[width=0.3\textwidth]{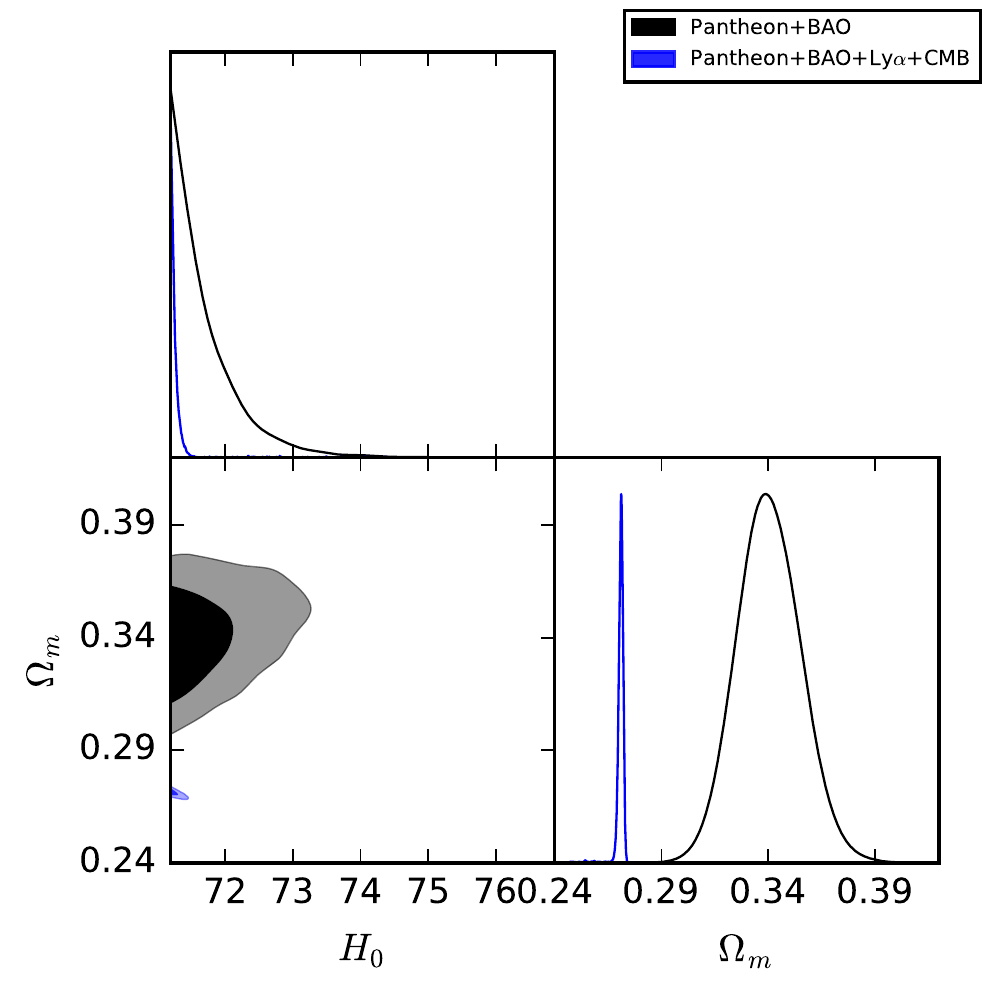}
\includegraphics[width=0.3\textwidth]{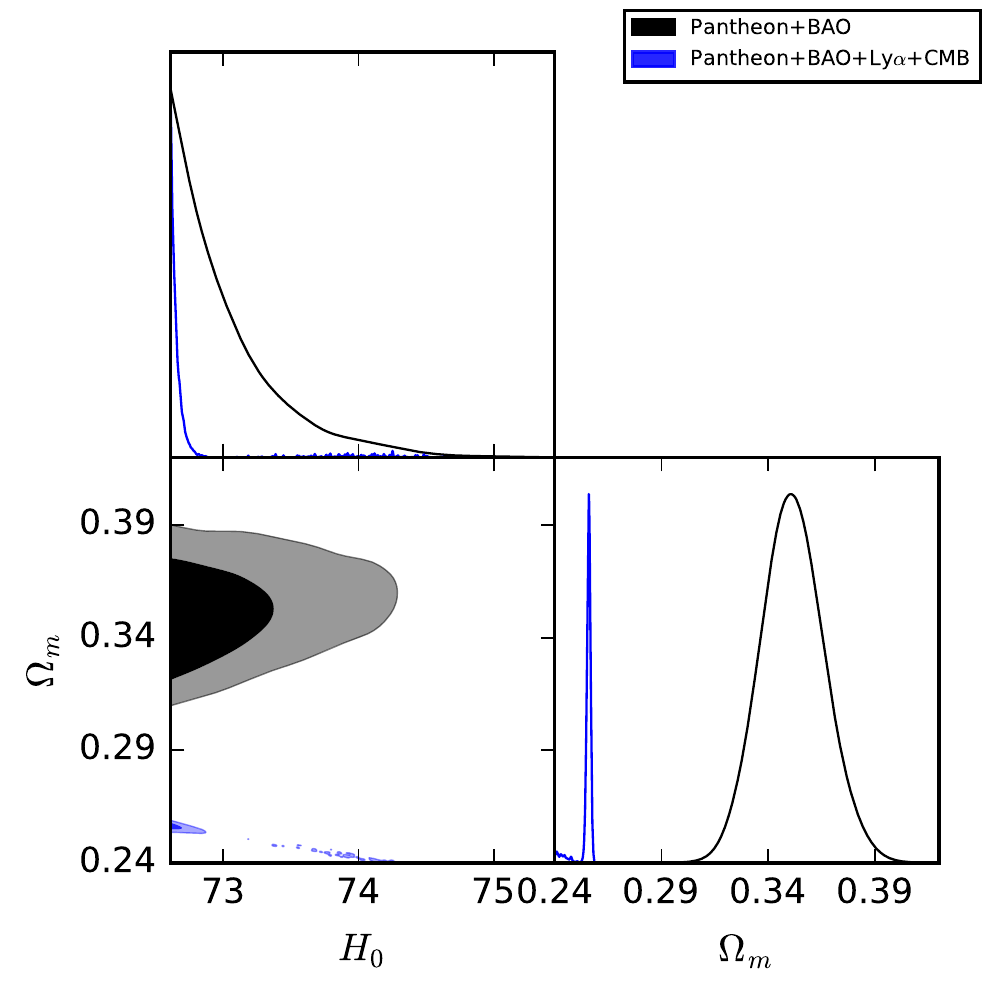}
\caption{2-D regions and 1-D marginalized distributions with 1$\sigma$ and 2$\sigma$ contours for $\LCDM$ model from different observations. From left to right, we use No $H_0$ prior, 2$\sigma$ hard-cut $H_0$ prior and 1$\sigma$ hard-cut $H_0$ prior from \cite{riess2019large}, respectively. The black curves/contours denote for the constraints from Pantheon+BAO and the blue ones are derived with Pantheon+BAO+Ly$\alpha$+CMB data combination.}
\label{fig:lcdm}
\end{figure*}

\begin{figure*}
\centering
\includegraphics[width=0.3\textwidth]{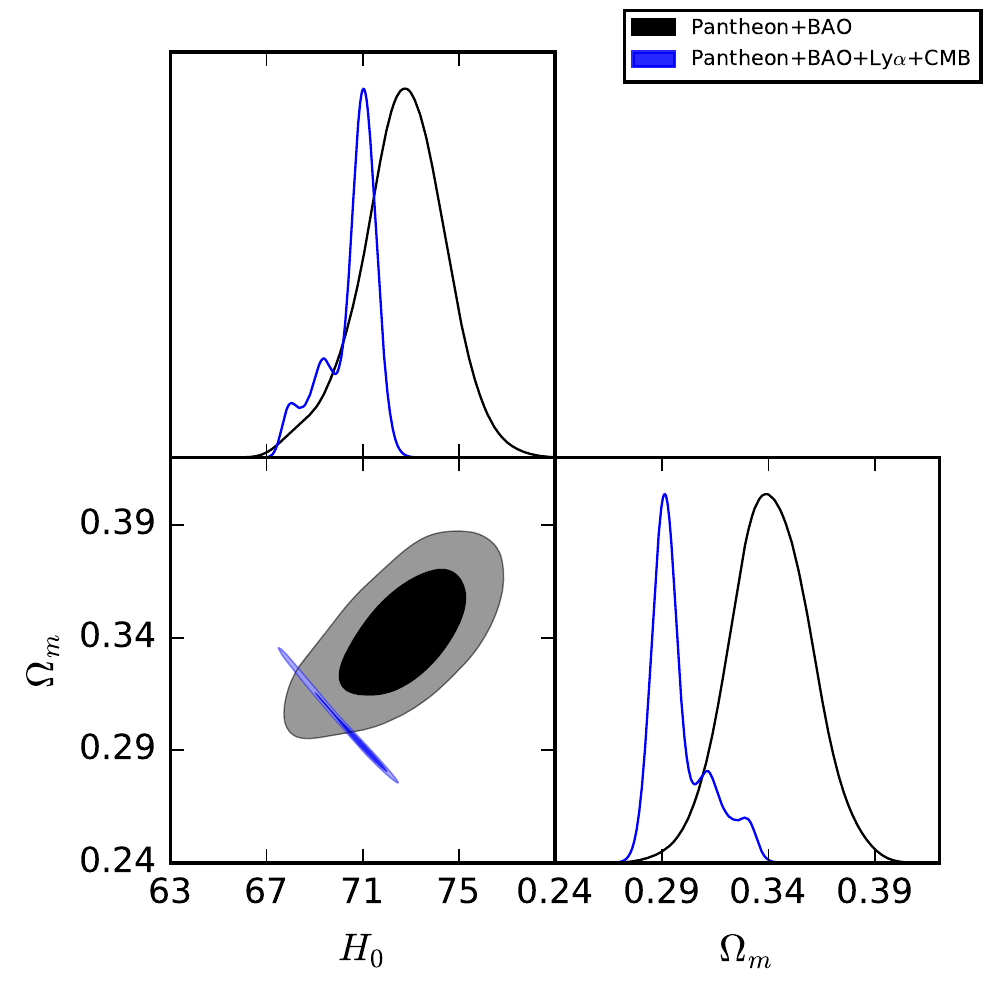}
\includegraphics[width=0.3\textwidth]{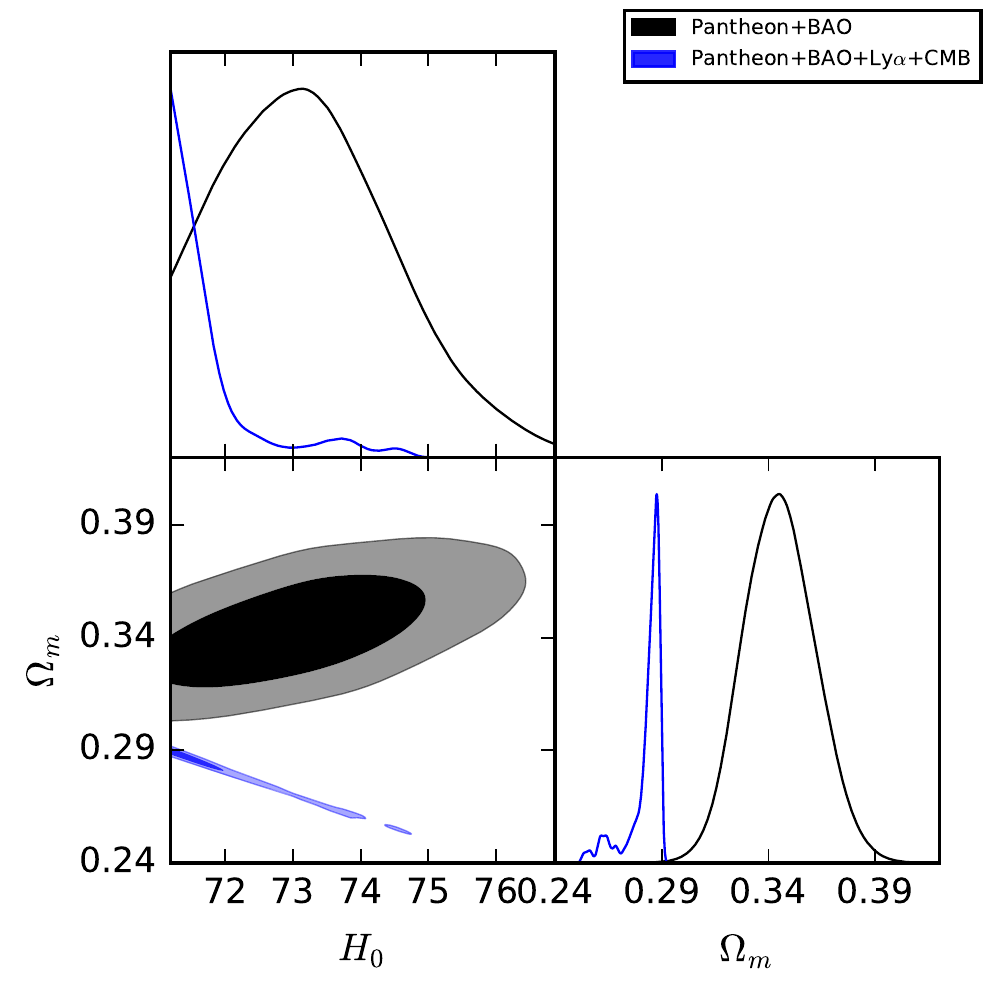}
\includegraphics[width=0.3\textwidth]{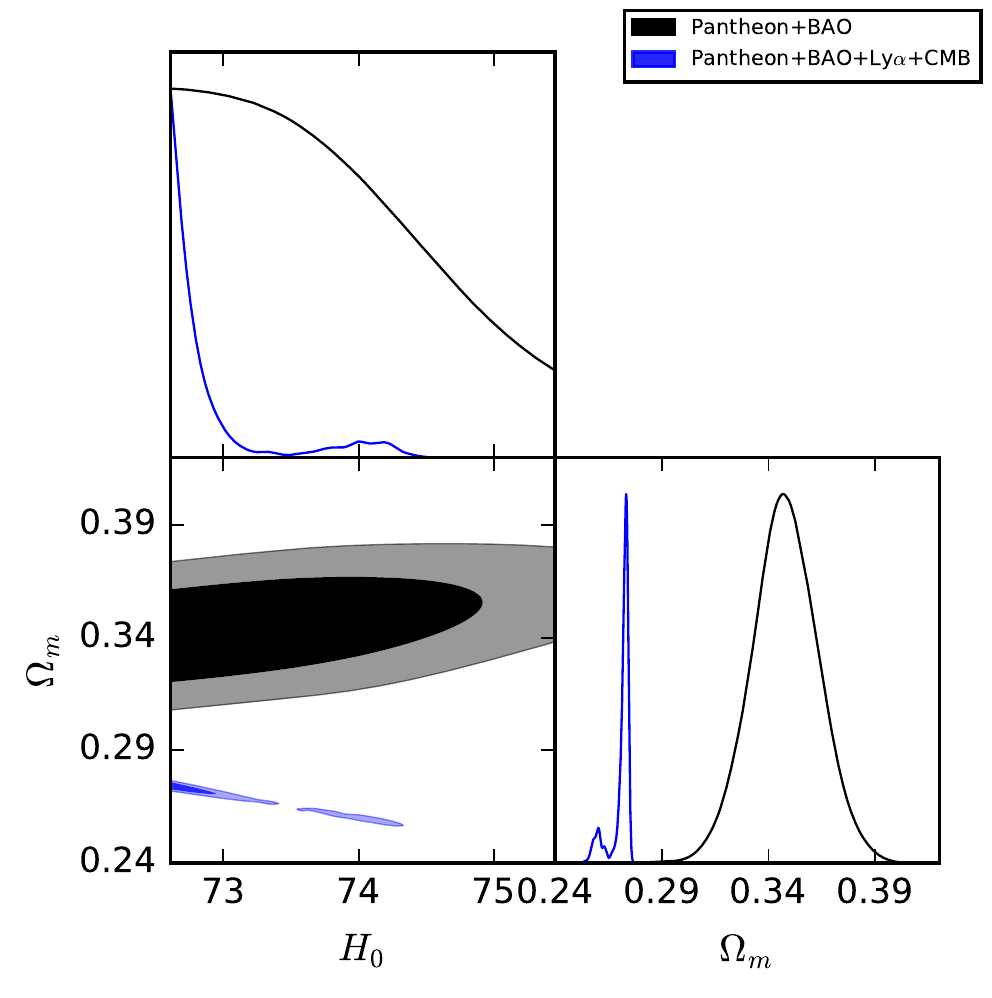}
\caption{2-D regions and 1-D marginalized distributions with 1$\sigma$ and 2$\sigma$ contours for PEDE model from different observations. From left to right, we use No $H_0$ prior, 2$\sigma$ hard-cut $H_0$ prior and 1$\sigma$ hard-cut $H_0$ prior from \cite{riess2019large}, respectively. The black curves/contours denote for the constraints from Pantheon+BAO and the blue ones are derived with Pantheon+BAO+Ly$\alpha$+CMB data combination.}
\label{fig:md2_tri}
\end{figure*}

\begin{table*}[!t]
\centering
\caption{Constraints on the parameters, $\chi^2_{bf}$ and the DIC for $\Lambda$CDM model, CPL parameterization and PEDE model are presented. Note that with hard-cut $H_0$ priors, the PEDE model is clearly outperforming $\Lambda$CDM model. With 1$\sigma$ hard-cut $H_0$ prior, the PEDE model is performing even better than CPL parameterization.} 
\label{tab:best_fit}
\begin{tabular}{c|c| c c c | c c c}
\hline
\hline 
  \multirow{2}{*}{Model} &  {Data}  & \multicolumn{3}{c|}{Pantheon+BAO}       & \multicolumn{3}{c}{Pantheon+BAO+Ly$\alpha$+CMB} \\
                         &  Parameters &  No $H_0$ Prior &   2$\sigma$ $H_0$ Prior  &  1$\sigma$ $H_0$ Prior   &  No $H_0$ Prior  &  2$\sigma$ $H_0$ Prior &   1$\sigma$ $H_0$ Prior     \\        
  \hline     
 \multirow{4}{*}{\bm{$\Lambda$}\textbf{CDM}} & {$\Om$}    & $0.299_{-0.043}^{+0.047}$   & $0.335_{-0.036}^{+0.040}$   &  $ 0.347_{-0.036}^{+0.041}$   & $0.311_{-0.014}^{+0.016}$ & $0.271_{-0.003}^{+0.002}$ &  $0.256_{-0.002}^{+0.002}$  \\
  \cline{2-8}
                          &{$H_0$}    & $66.94_{-3.256}^{+3.721}$   & $71.19_{0.0}^{+1.890}$  &  $72.61_{-0.000}^{+1.617}$ &  $67.91_{-1.150}^{+1.074}$ & $71.19_{-0.000}^{+0.271}$ &    $72.61_{-0.000}^{+0.200}$  \\
  \cline{2-8}
                          &\bm{$\chi^2_{bf}$}  & {1046.94}  & \textbf{1054.76}  & \textbf{1060.25} & {1056.12} &\textbf{1112.28}  & \textbf{1168.98}  \\
  \cline{2-8}
                          &\textbf{DIC}        & {1051.00}  & \textbf{1058.88}  & \textbf{1064.27} &  {1062.35}  & \textbf{1127.03} & \textbf{1195.07}  \\
\hline
\hline
 \multirow{6}{*}{\textbf{CPL}} &{$\Om$}    & $0.285_{-0.180}^{+0.113}$ & $0.332_{-0.050}^{+0.071}$  & $0.350_{-0.043}^{+0.050}$   & $0.307_{-0.021}^{+0.026}$  & $0.286_{-0.011}^{+0.007}$ &  $0.274_{-0.009}^{+0.006}$ \\
          \cline{2-8}
  &  {$H_0$}             & $64.84_{-16.12}^{+14.49}$ & $71.30_{-0.117}^{+5.561}$ & $72.70_{-0.091}^{+2.746}$  & $ 68.49_{-2.680}^{+2.302}$ & $71.19_{-0.002}^{+1.277}$ &  $72.61_{-0.004}^{+0.918}$ \\
  \cline{2-8}
          &{$w_0$}           & $-0.82_{-0.541}^{+0.193}$  & $-1.08_{-0.347}^{+0.422}$  & $-1.05_{-0.347}^{+0.350}$ & $ -0.98_{-0.218}^{+0.267}$ & $-1.07_{-0.240}^{+0.259}$ & $ -1.13_{-0.206}^{+0.274}$ \\
          \cline{2-8}
          & {$w_a$}         & $0.675_{-3.103}^{+0.547}$ & $-0.11_{-3.192}^{+1.510}$ & $-0.46_{-2.686}^{+1.830}$   & $ -0.16_{-1.109}^{+0.816}$ & $-0.20_{-1.240}^{+0.986}$ & $ -0.11_{-1.321}^{+0.728}$  \\
          \cline{2-8}
          & \bm{$\chi^2_{bf}$}            & {1044.98} & \textbf{1048.84}  &\textbf{1049.66}   & {1055.52} &  \textbf{1066.85} & \textbf{1080.83}\\
        \cline{2-8}
        &\textbf{DIC}      & {1052.59} & \textbf{1054.46} & \textbf{1056.23}  & {1065.48} & \textbf{1085.06} &\textbf{1128.50}    \\
\hline
\hline
 \multirow{4}{*}{\textbf{PEDE}}         &  {$\Om$}   & $0.341_{-0.041}^{+0.045}$  & $0.341_{-0.037}^{+0.041}$ & $0.341_{-0.030}^{+0.041}$ & $0.291_{-0.016}^{+0.015}$  & $0.289_{-0.014}^{+0.002}$ & $0.274_{-0.006}^{+0.002}$\\
           \cline{2-8}
        &   {$H_0$}   & $72.84_{-3.530}^{+3.814}$ & $73.01_{-1.8231}^{+3.371}$  &$72.79_{-0.186}^{+2.652}$    & $71.02_{-1.368}^{+1.452}$ & $71.19_{-0.001}^{+1.306}$ &   $72.61_{-0.000}^{+0.651}$ \\
          \cline{2-8}
          & \bm{$\chi^2_{bf}$}    & {1050.04} & \textbf{1050.04} &\textbf{1050.04}  & {1071.12}& \textbf{1071.20} & \textbf{1080.40} \\
        \cline{2-8}
         & \textbf{DIC}           &   {1052.01}  & \textbf{1053.33} &  \textbf{1052.98} & {1091.15} & \textbf{1091.65} & \textbf{1100.94}  \\
\hline
\hline
\end{tabular}
\end{table*}

\begin{figure*}
\centering
\includegraphics[width=0.45\textwidth]{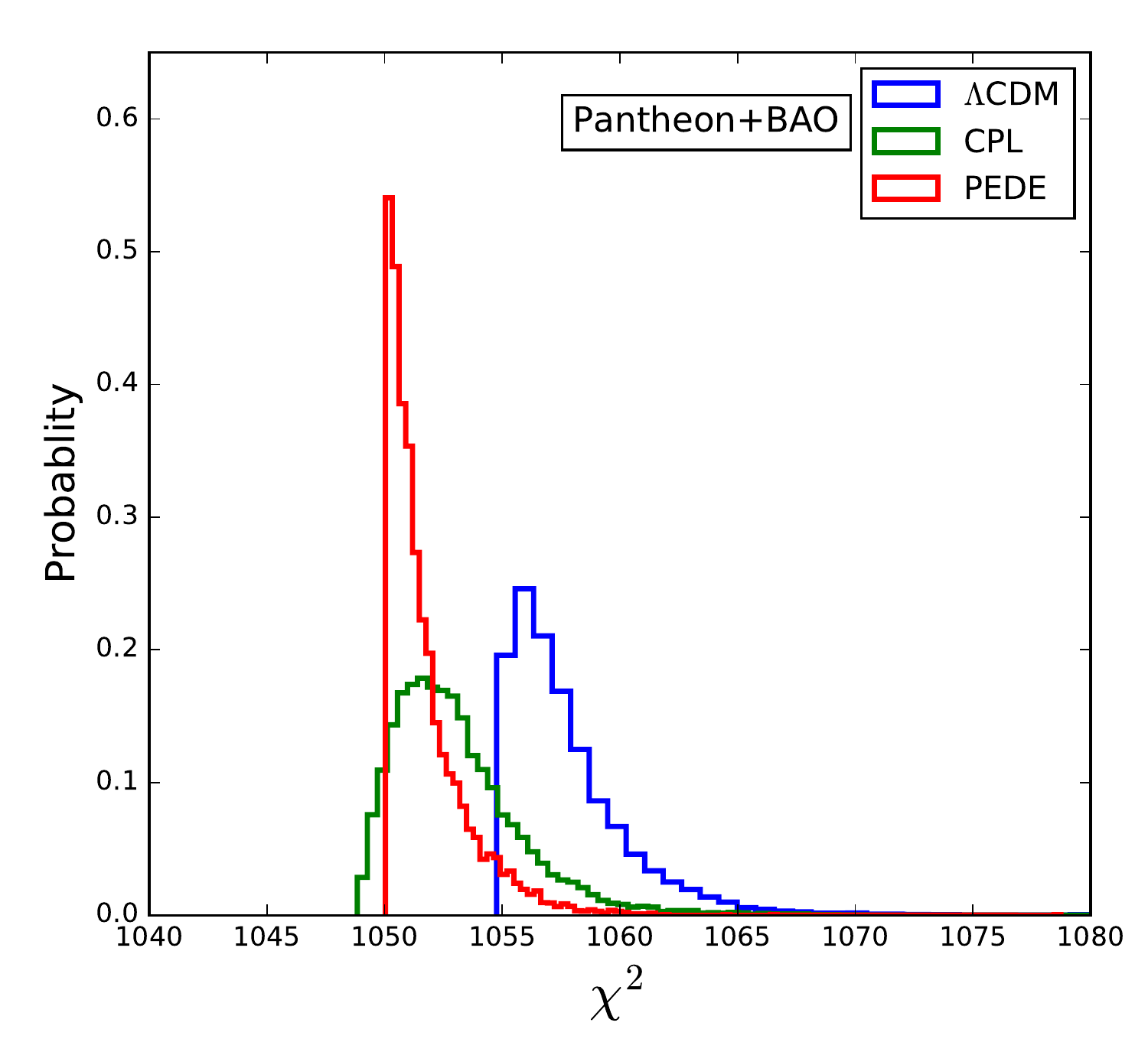}
\includegraphics[width=0.45\textwidth]{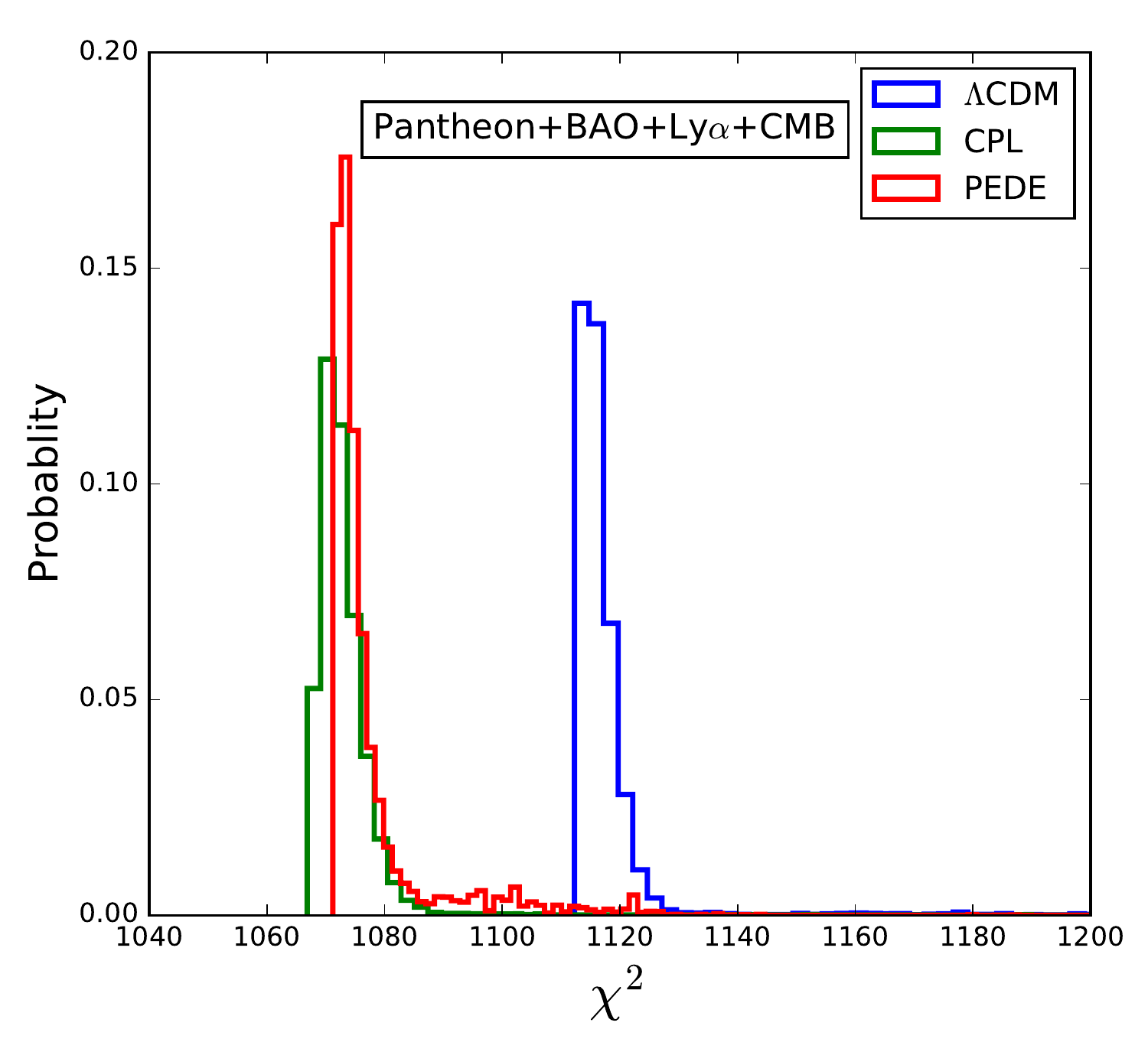}
\includegraphics[width=0.45\textwidth]{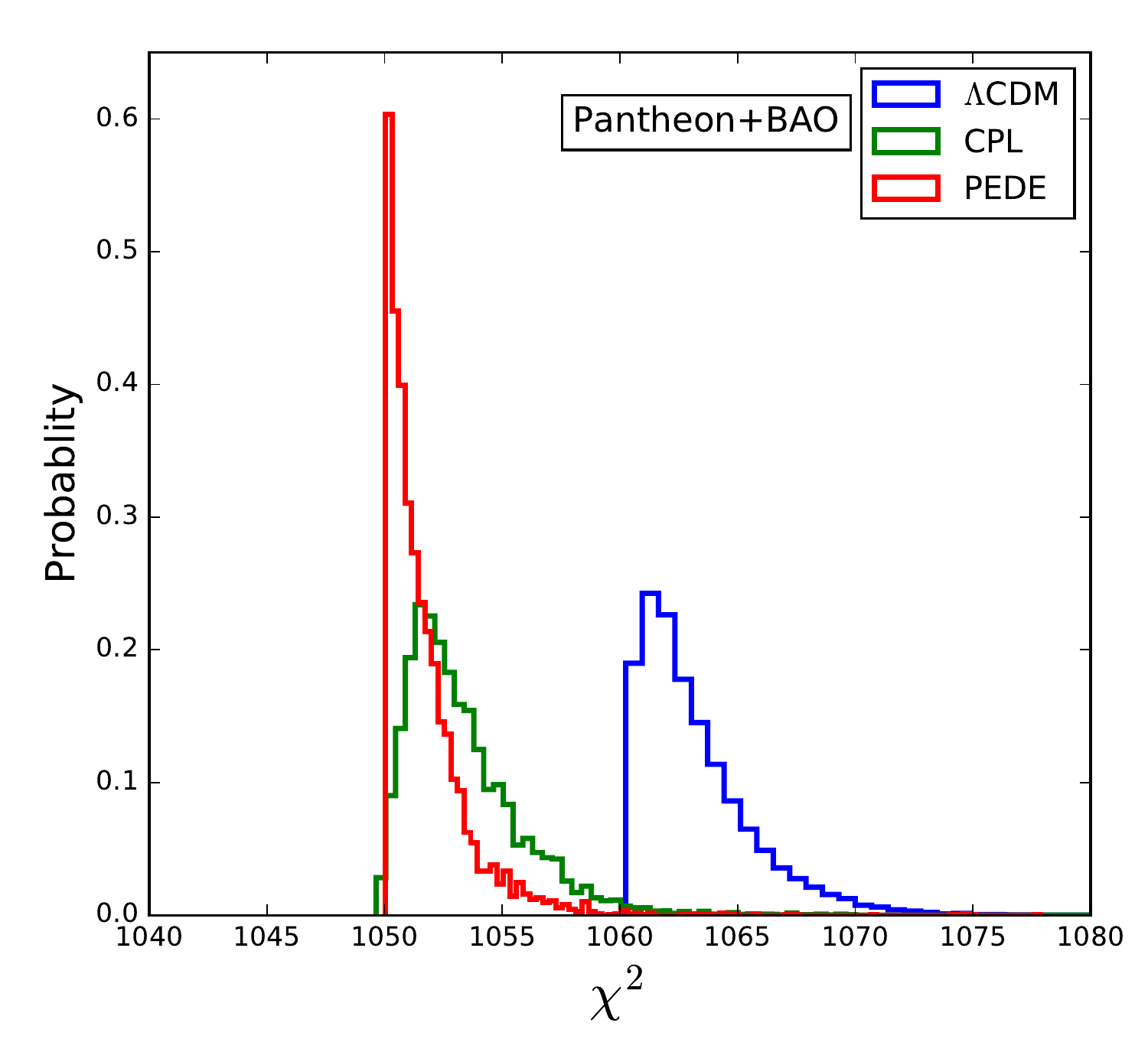}
\includegraphics[width=0.45\textwidth]{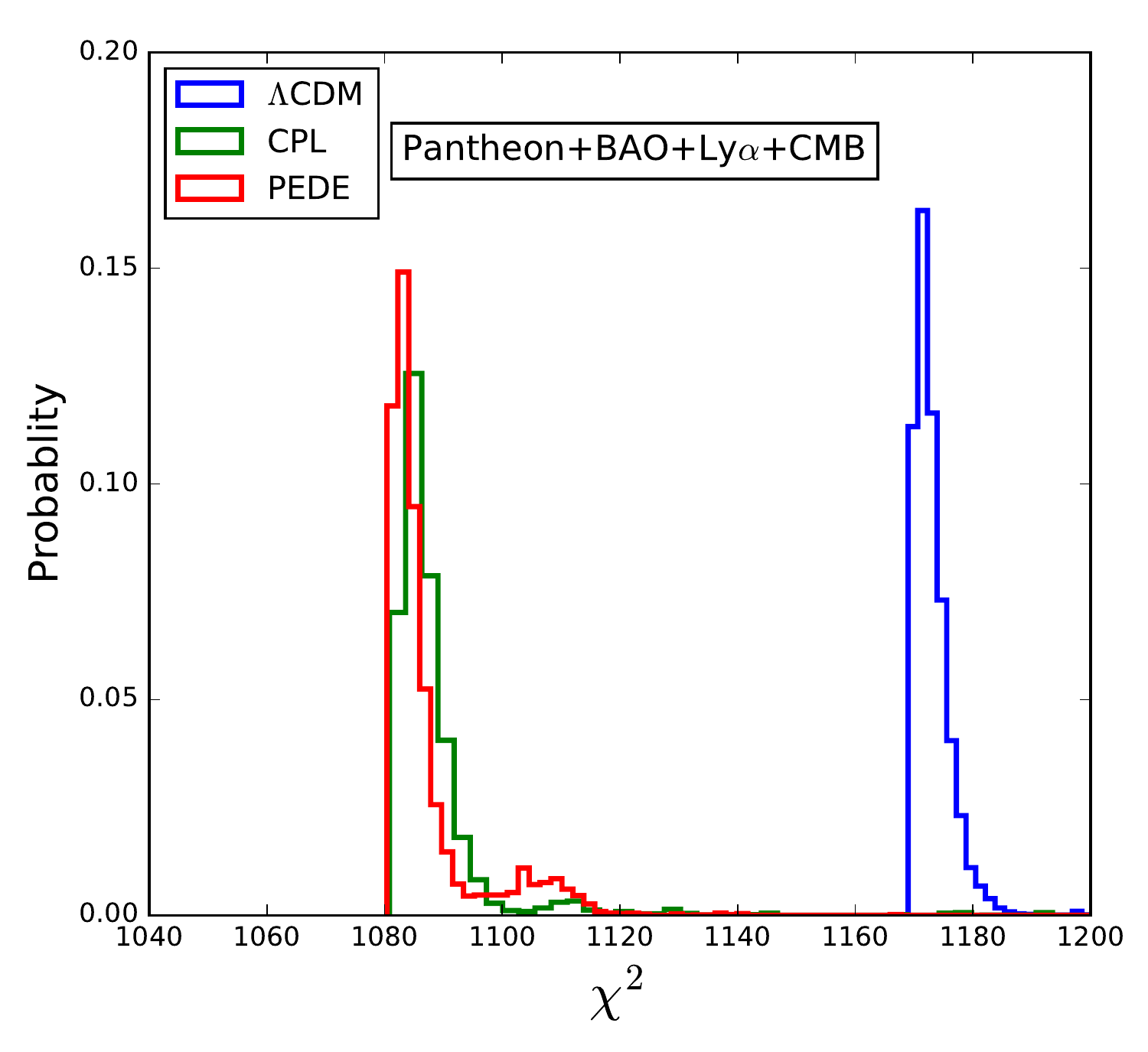}
\caption{The histograms of $\chi^2$ distribution from the converged MCMC chains for $\Lambda$CDM model, CPL and PEDE are presented. The left plots shows the $\chi^2$ distribution for the {Pantheon+BAO} combination and the right plots are obtained with {Pantheon+BAO+Ly$\alpha$+CMB} combination. Upper plots are derived with setting 2$\sigma$ $H_0$ hard-cut prior and lower plots are derived with setting 1$\sigma$ hard-cut $H_0$ prior. Combining all the data, there is hardly an overlap between the $\chi^2$ distribution of the PEDE model and $\Lambda$CDM model that explains the huge difference we derived for their DIC.}
\label{fig:chi2_2H0p}
\end{figure*}

We show the results for $\LCDM$ in Fig.~\ref{fig:lcdm}, in which we present the 2D regions and 1D marginalized distributions with 1$\sigma$ and 2$\sigma$ contours from different data combinations. The left panel shows the results with No $H_0$ prior, and the middle and right panels show the results of setting hard-cut 2$\sigma$ $H_0$ prior and 1$\sigma$ $H_0$ prior, respectively.
 Fig.~\ref{fig:md2_tri} shows the results for our PEDE model. Comparing Fig.~\ref{fig:lcdm} and Fig.~\ref{fig:md2_tri}, we can find that PEDE model pushes the values of both $H_0$ and $\Om$ toward a higher direction for Pantheon+BAO data sets when No $H_0$ prior is considered. However adding CMB and high redshift BAO measurements makes the constraints on value of $\Om$ slightly smaller. While the tension in estimated value of the Hubble constant is relieved in PEDE model, some tension in estimated value of the matter density persist (though substantially reduced in comparison with the case of $\Lambda$CDM model).

The parameter constraints for $\LCDM$ model, $w_0$-$w_a$ parameterization and PEDE model are summarized in Table~\ref{tab:best_fit}, in which we also show the best fit $\chi^2$ and DIC values for each model from different data combinations. The $\chi^2$ distributions for the converged MCMC chains for the $\LCDM$ model, $w_0$-$w_a$ parameterization and PEDE model from lower redshift observations (left) and combined observations (right) are shown in Fig.~\ref{fig:chi2_2H0p}. The upper plots are based on a hard-cut 2$\sigma$ $H_0$ prior and the lower plots are based on a hard-cut 1$\sigma$ $H_0$ prior. From Table~\ref{tab:best_fit} and Fig.~\ref{fig:chi2_2H0p} we can see that, PEDE model provides with substantially better $\chi^2_{bf}$ with respect to $\LCDM$ model considering 2$\sigma$ $H_0$ prior, with $\Delta \chi^2_{bf}\,=\,-4.72$ for lower redshift observations and $\Delta \chi^2_{bf}\,=\,-41.08$ for the combined observations. When calculating DIC for different models, we find $\Delta\,{\rm{DIC}}\,=\,-5.55$ and $\Delta\,{\rm{DIC}}\,=\,-35.38$ with respect to $\LCDM$ model for lower redshifts and combined observations, respectively. DIC for PEDE model is very much comparable with $w_0$-$w_a$ parameterization when setting 2$\sigma$ $H_0$ prior.

As can be seen from Table~\ref{tab:best_fit} and lower plots in Fig.~\ref{fig:chi2_2H0p}, with 1$\sigma$ hard-cut $H_0$ prior, the $\chi^2_{bf}$ of PEDE model becomes much lower than that of $\LCDM$ model, with $\Delta \chi^2_{bf}\,=\,-10.21$ for lower redshift observations and $\Delta \chi^2_{bf}\,=\,-88.58$ for combined observations. This is comparable to $w_0$-$w_a$ parameterization model, which has 2 more degree of freedom. When calculating DIC values, PEDE model gives best results among the three models we considered, with $\Delta\,{\rm{DIC}}\,=\,-94.13$ with respect to $\LCDM$ model and $\Delta\,{\rm{DIC}}\,=\,-27.56$ with respect to $w_0$-$w_a$ parameterization for combined observations. Lower plots in Fig.~\ref{fig:chi2_2H0p} clearly shows how the proposed PEDE model outperforms $\Lambda$CDM model if we set hard-cut $H_0$ priors and effectively ruling it out with high statistical significance where the tail of $\chi^2$ distribution for this model has no overlap with the same distribution for the case of $\Lambda$CDM model. We should note that the considered 2$\sigma$ and 1$\sigma$ hard-cut priors for the Hubble constant that effectively affects the assumed models from the lower bound, associate to $97.72\%$ and $84.13\%$ probabilities respectively. In other words there is $97.72\%$ chance that our results for 2$\sigma$ $H_0$ prior holds with future observations (with higher precision) and there is $84.13\%$ chance that our results with 1$\sigma$ $H_0$ prior holds with future high precision observations.

\section{Conclusion} \label{sec:con}
We propose a simple phenomenologically emergent model of dark energy that has zero degrees of freedom, which is similar to the case of cosmological constant. The proposed functional form based on a hyperbolic tangent function has a symmetrical behavior in dark energy density as a function of the scale factor in logarithmic scales. The argument behind having the pivot of symmetry at current time can be associated with the fact that dark energy and matter densities are comparable at the current time. This model can be trivially modified to set the pivot of symmetry at the scale of dark energy-dark matter density equality which would be at $z\approx 0.3$. In our proposed PEDE model, dark energy has no effective presence in the past and its density increases to double of its current value in the far future. Theoretically this will be associated with a dark energy component with $w\,=\,-\frac{2}{3{\rm{ln}}\,10}-1$ in the past that will evolve to $w\,=\,-1$ in the far future. 

Setting hard-cut 2$\sigma$ and 1$\sigma$ priors on Hubble constant from local measurements, associated with $97.72\%$ and $84.13\%$ probabilities respectively, and using most recent cosmological observations from low and high redshift universe, our proposed model surpasses cosmological constant with large margins. Assuming reliability of the Hubble constant measurement and no substantial systematic in any of the data we used, with 2$\sigma$ and 1$\sigma$ hard-cut priors of $H_0$ our proposed PEDE model rules out cosmological constant with large statistical significance with $\Delta\,{\rm{DIC}}\,=\,-35.38$ and $\Delta\,{\rm{DIC}}\,=\,-94.13$ respectively. It is indeed interesting that with 1$\sigma$ hard-cut prior on $H_0$, this model can even outperform the widely used $w_0$-$w_a$ parametric form with $\Delta\,{\rm{DIC}}\,=\,-27.56$. This can be a game changer as our proposed model can establish itself as an strong alternative and favorite to the cosmological constant in the current standard model of cosmology. 

With no information on the Hubble constant, the concordance $\LCDM$ model seems to be the most favored model. Consequently, all our results and the conclusion on ruling out $\Lambda$ is solely and directly associated with the reliability of the Hubble constant measurement. 

While our proposed model can significantly reduce the tensions in estimation of the cosmological parameters using low- and high- redshift data, some level of tension remains, in particular in the estimation of the matter density. This matter requires further study in order to understand the origin of any discrepancy that persist in any model assumption. More detailed studies are required to compare our proposed model to different cosmological observations that can have some traces of $\LCDM$ assumptions in their pipelines. However, it is evident that making more appropriate treatment of different data for our proposed model can only make this model to perform better with respect to $\LCDM$ model. 

Assuming that current cosmological data are all viable, our proposed model is shown to be a better representative of the effective behavior of dark energy in comparison with the cosmological constant. This work can guide theoretical studies of dark energy and our Universe in general. 

We should recall that our ultimate goal should be to find a theoretical explanation for dark energy or, in a more fundamental approach, for the whole dark sector considering both dark matter and dark energy. We should consider different possibilities and look for the correct theory of gravity; considering dark energy and dark matter as curvature effects or unifying the whole dark sector might be reasonable ways to explain theoretically the observationally supported emergent behavior of the effective dark energy~ \citep{capozziello2006dark,yang2019dawn}. Distinguishing between physical and geometrical models of dark energy as well as modified theories of gravity and breaking the degeneracies is in fact a fundamental task in cosmology that might be achievable by cosmography~\citep{2013PhRvD..87b3520S,Shafieloo:2018gin,Capozziello:2019cav}. Future observations would shed light on this important problem.

We should note that at the latest stages of this work, we became aware of the work of \citet{keeley2019implications}, which discussed a similar behavior of dark energy, but employed a parametric form that has few degrees of freedom similar to what has been introduced earlier in \citet{bassett2002late,shafieloo2009cosmic}. The simplicity of our phenomenological model with zero degrees of freedom for dark energy sector and its great performance is the core of our analysis which allow us to rule out cosmological constant with large statistical significance when we set hard priors on the Hubble constant.    

\acknowledgments
X. Li and A.S. would like to acknowledge the support of the National Research Foundation of Korea (NRF-2016R1C1B2016478). X. Li is supported by the Strategic Priority Research Program of the Chinese Academy of Sciences, Grant No. XDB23000000. A.S. would like to acknowledge the support of the Korea Institute for Advanced Study (KIAS) grant funded by the Korea government. This work benefits from the high performance computing clusters Polaris and Seondeok at the Korea Astronomy and Space Science Institute. A.L. and A.S. would like to thank Zong-Hong Zhu and Beijing Normal University for the hospitality during the early stages of this work.

\bibliography{references}
\end{document}